\documentclass[useAMS,usenatbib,usegraphicx]{mn2e}
\usepackage{epsfig}
\usepackage{amsmath} 
\usepackage{rotating}           
\usepackage{color}                

\usepackage{times}

\def\kms{km ${\rm s}^{-1}$}

\def\ch2{$\chi^2$}

\def\kms {\hbox{${\rm km\ s}^{-1}$}}

\def\scm  {$\hbox{{\rm cm}}^{-2}$}    

\def \AL {$\alpha $}     
\def \HI {H{\sc \,i}}

\def\lapp{\ifmmode\stackrel{<}{_{\sim}}\else$\stackrel{<}{_{\sim}}$\fi}
\def\gapp{\ifmmode\stackrel{>}{_{\sim}}\else$\stackrel{>}{_{\sim}}$\fi}

\usepackage{xspace}
\newcommand{\oiii}{\hbox{[O\,\textsc{iii}]}\xspace}

\title[A third 21-cm absorber towards MG J0414+0534]{A third \HI\ 21-cm absorption system in the sight-line of MG~J0414+0534: A redshift for Object X?}
\author[S. J. Curran et al.]{S. J. Curran$^{1}$\thanks{E-mail:
sjc@phys.unsw.edu.au}, M. T. Whiting$^{2}$,  A. Tanna$^{1}$, C. Bignell$^{3}$ and J. K. Webb$^{1}$\\
$^{1}$School of Physics, University of New South Wales, Sydney NSW 2052, Australia\\
$^{2}$CSIRO Australia Telescope National Facility, PO Box 76, Epping NSW 1710, Australia\\
$^{3}$National Radio Astronomy Observatory, P.O. Box 2, Rt. 28/92 Green Bank, WV 24944-0002, USA}

\begin{document}

\date{Accepted ---. Received ---; in original form ---}

\pagerange{\pageref{firstpage}--\pageref{lastpage}} \pubyear{2011}

\maketitle

\label{firstpage}

\begin{abstract}
  We report the detection of a third \HI\ 21-cm absorber in the sight-line towards the $z=2.64$ quasar MG~J0414+0534
  (4C\,+05.19).  In addition to the absorption at the host redshift and in the $z=0.96$ gravitational lens, we find,
  through a decimetre-wave spectral scan towards this source, strong absorption at $z=0.38$. We believe this may be
  associated with ``Object X'', an additional feature apparent in the field of the lensing galaxy and lensed images, on
  the basis of its close proximity to the quasar images and the possible detection of the \oiii\ doublet in a published
  optical spectrum.  If real, the strength of the \oiii\ emission would suggest the presence of an active galactic
  nucleus, or a gas-rich galaxy undergoing rapid star formation, either of which is consistent with the strong outflows
  apparent in the 21-cm spectrum. Although this is the strongest intervening 21-cm absorber yet found (a column density
  of $N_{\rm HI} \gapp10^{22}$ \scm, for a modest $T_{\rm s}/f\gapp300$~K), simultaneous observations failed to detect
  any of the 18-cm OH lines at the 21-cm redshift. This suggests that, as for the lensing galaxy, this is not the
  primary location of the intervening material responsible for the very red colour of MG~J0414+0534.
\end{abstract}


\section{Introduction}
\label{intro}

Radio-band observations of absorption systems along the sight-lines to distant quasars provide a 
powerful probe of the cool atomic and molecular gas at high redshift. This gas constitutes the reservoir
of raw material which forms stars, planets and all other non-diffuse structures in the early Universe.
As well as giving insight into how the contents of present day galaxies came to be, redshifted \HI\ 21-cm and 
OH 18-cm absorption lines have the potential to be very useful in determining whether the fundamental
constants of nature have changed since these large look-back times (see \citealt{cdk04} and references therein).

Unfortunately, such absorption is currently rare, with only 76 \HI\ 21-cm absorbers known at
$z\geq0.1$, 41 of which are due to intervening systems (summarised in \citealt{cur09a})\footnote{With the
  addition of one new intervening absorber reported in \citet{sgp+10}.}, with 35 being associated
with the quasar/quasar host providing the background illumination (summarised in
\citealt{cw10})\footnote{With the addition of three new associated absorbers, two reported in
  \citet{cwm+10} and one in \citet{cwwa11}.}.  OH 18-cm absorption is rarer still, with only five
systems known \citep{cdn99,kc02a,kcdn03,kcl+05}, three of which are intervening and two being
associated with the background source. 

%

All of the OH and 80\% of detected \HI\ absorption occurs at $z\lapp1$. Much of this bias is due to 
the limited availability of interference free
bands at low frequencies, although there are additional effects contributing to lower detection rates at high redshift:
For the intervening, systems the 21-cm
detection rates  (61\% at $z\lapp1$ cf. 33\% at $z\gapp1$, \citealt{cur09a}) can be attributed to the geometry effects introduced by a flat
expanding Universe, causing the coverage of the background flux to be systematically lower at higher redshift
\citep{cw06,ctd+09}. For associated systems, the rates (39\% at $z\lapp1$ cf. 17\% at $z\gapp1$)  are biased by the
traditional optical selection of targets, where only the most ultra-violet luminous sources
are known at high redshift, since the intense UV flux from the near-by active galactic nucleus ionises/excites
the cool gas beyond detection \citep{cww+08}. Although both of these effects are present in some cases at $z\lapp1$, although
they are always present for the high redshift sources.

Optical selection effects further compound the detection of OH absorption in that, despite much searching of
objects in the millimetre-band,  where four of the five OH absorbers were originally discovered, millimetre-wave
absorption has yet to be found in an optically selected target (see \citealt{cmpw03}). \citet{cwm+06} suggest
that this is due to the optical brightness of these objects selecting against the dustier, and thus most 
molecular friendly absorbers. This is demonstrated through the optically selected damped Lyman-$\alpha$ absorption systems (DLAs) 
having optical--near infrared colours of $V-K\lapp4$ and
molecular fractions of ${\cal F}\equiv\frac{2N_{\rm H_2}}{2N_{\rm H_2}+N_{\rm HI}}\sim10^{-7} - 0.3$, whereas the 
radio absorbers have $V-K\gapp5$ and ${\cal F}\approx0.6 - 1$ (see
figure 3 of \citealt{cww09}). 

This is strong evidence that the background quasar light is reddened by the dust in the foreground
absorber, which protects the molecular gas from the harsh UV environment. Thus, in order to increase
the number of redshifted OH (and \HI) absorbers known, we should target the reddest objects\footnote{This prediction
was verified through the detection of OH at $z=0.76$ in the gravitational lens intervening PKS 0132--097 \citep{kcl+05}:
With the detection of 21-cm absorption \citep{kb03} and an optical--near-infrared colour of $V-K =
  8.92$ along this sight-line \citep{glw+02}, this was a prime  target.
  On this basis, we ourselves attempted to detect OH with both the Giant Metrewave Radio Telescope (GMRT) and the
  Westerbork Synthesis Radio Telescope in February 2005. However, severe RFI at 945 MHz ruined both
  observations.}.
However, due to their very faintness, optical spectra are not generally available and
so we do not have a redshift to which to tune the telescope. We have therefore embarked on a 
programme of wide-band (200 \& 800 MHz) spectral scans of very red ($V-K\gapp6$) radio-loud
objects with the Green Bank Telescope (GBT) in search of the dust and molecular gas responsible for
the obscuration. 

In this letter we report the detection of very strong 21-cm absorption at $z=0.38$ towards the 
$z= 2.64$ quasar MG J0414+0534 (4C\,+05.19), where we have previously detected 21-cm in the $z=0.96$
gravitational lens \citep{cdbw07}. With an optical--near-infrared colour of $V-K = 10.26$, this is the
reddest of our targets, and, although 21-cm has also been detected in the lens, as well as in the
host galaxy \citep{mcm98}, OH remains undetected at either of these three redshifts.

\section{Target selection and observations}

As mentioned above, all of our targets were selected on the basis of their large
optical--near-infrared colours ($V-K\gapp6$) and high radio fluxes ($\gapp1$ Jy).  Being a database
of bright and generally compact objects, with comprehensive optical photometry \citep{fww00},
these were taken from the Parkes Half-Jansky Flat-spectrum Sample (PHFS, \citealt{dwf+97}), which,
with the above conditions, gave five sources which could be scanned for both \HI\ and OH by the GBT. However, given the
very wide band-widths required for the full spectral scans, much of the data are subject to
severe radio frequency interference (RFI), with only limited parts of the band being useful (Tanna et al., in prep). 
Nevertheless, we are able to obtain enough useful data on MG
J0414+0534 to reveal a clear, strong detection of neutral
hydrogen. 

The 0.91--1.23 GHz range of the J0414+0534 observations were performed on 23 January 2007 using the PF2
receiver and the autocorrelation spectrometer over a 200 MHz wide bandpass (with 16\,384 lags), centred on 1.0 GHz in
two  orthogonal linear polarisations.  This band was observed for a total of two hours in 5~min position-switched scans, and 
the removal RFI affected scans left 44 minutes of good data, with a
mean system temperature of $T_{\rm sys} = 24$~K and an r.m.s. noise level of $\approx10$ mJy per 3.56 \kms\ (at 1030 MHz) channel in the
clear parts of the bandpass.

\begin{figure}
\centering \includegraphics[angle=270,scale=0.60]{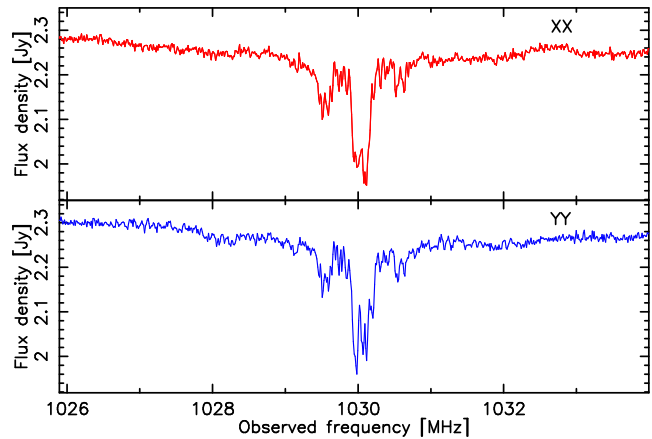}
\caption{The absorption profile at $1030$ MHz  towards J0414+0534 in each of the two orthogonal linear polarisations. The 
data are shown at the observed 12.207 ~kHz channel spacing, which gives 3.56 \kms\ at 1030 MHz. The flux density is
found to be 2.21 Jy at 1.03 GHz, cf. the 2.12 Jy at 1.4 GHz \citep{kmh97}.}
\label{pols}
\end{figure}
The data were calibrated, flagged and averaged using the {\sc gbtidl} package and, as seen in each of the polarisations, an absorption
feature was detected close to 1030 MHz (Fig. \ref{pols}). Due to the structure of the line
and the strength of the main component (a velocity integrated optical depth of $\int\!\tau\,dv \approx
11$ \kms), we believe that the feature is due to 21-cm absorption at a redshift of $z \approx0.38$.

\section{Results and discussion}
\label{randd}
\subsection{The \HI\ 21-cm absorption}
\label{hi}

\begin{figure}
\centering \includegraphics[angle=0,scale=0.35]{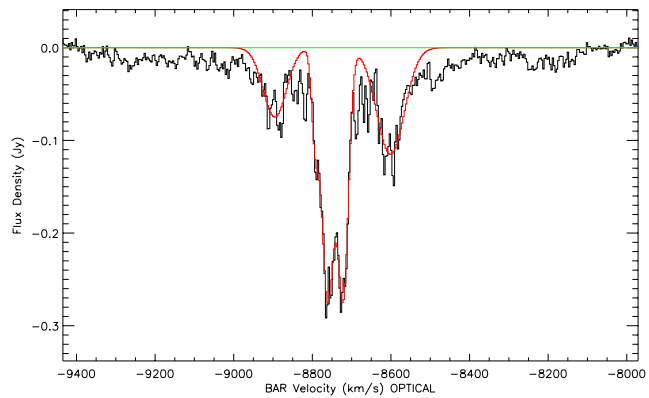} 
\caption{Gaussian fits to the absorption profile, where we fit three
Gaussians to the central feature and one to each of the outlying features
(summarised in Table \ref{fits}). The flux scale is relative to 2.21 Jy and the velocity offset is
relative to the central observing frequency of 1.000 GHz.  As Fig.~\ref{pols}, the 
velocity resolution is 3.56~\kms.}
\label{gauss}
\end{figure}

In each polarisation (Fig. \ref{pols}) it is clear that the absorption is comprised of several components and in Fig. \ref{gauss}
we show the Gaussian fits to the profile and summarise these in Table \ref{fits}. 
\begin{table*}
 \centering
 \begin{minipage}{170mm}
   \caption{Gaussian fits to the \HI\ absorption properties at $z=0.38$ towards MG J0414+0534 derived from the Gaussian
     fits (in the order left to right shown in Fig. \ref{gauss}).  $\Delta v$ gives the line offset from the reference
     1000.02 MHz, 
     $\nu$ [MHz] is the observed frequency (barycentric, optical definition), $z_{\rm abs}$ the corresponding redshift,
     FWHM is the full-width half maximum of the line [\kms], followed by the observed peak depth [mJy], which
 gives the observed optical depth, which in the optically thin regime is
     related to the actual optical depth via $\tau=f.\tau_{\rm act}$, where $f$ the covering factor of the background
     continuum source.  The last two columns give the velocity integrated depth [\kms] and the derived column density
     [\scm], where $T_{\rm s}$ is the spin temperature [K].  \label{fits}}
  \begin{tabular}{@{} c c c c c c cc c@{}}
  \hline
Line & $\Delta v$ & $\nu$  & $z_{\rm abs}$ &  FWHM  & Depth  &$\tau$ &$\int\!\tau\,dv$ & $N_{\rm HI}$ [$\times10^{18}.\,(T_{\rm s}/f)$] \\
  \hline
1 &   --8895    &   $1030.57\pm 0.03 $  &  $0.37827\pm0.00004$   &  $70\pm10$    &  $-75\pm6$ &   $0.033\pm0.002$   &  $2.3\pm0.5$   &   $4.2\pm0.9$   \\
2 &   --8793   &   $1030.216\pm0.008$  &  $0.37875\pm0.00001$        &     $21\pm5$       & $-90\pm2$ &     $0.041\pm0.001$     &   $0.9\pm0.2$   &  $1.9\pm0.4$  \\
3 &  --8760    &   $1030.100\pm 0.004$ &   $0.378901\pm0.000005$     &  $36\pm6$         &  $-274\pm9$ &    $0.123\pm0.004$    &   $4.4\pm0.9$  & $8.0\pm1.6$\\
4 &   --8721    &   $1029.962\pm 0.005$ & $0.379086\pm0.000007$   &  $31\pm3$        &  $-263\pm14$ &   $0.119\pm0.006$    &    $3.7\pm0.5$  &   $6.7\pm0.9$ \\ 
5 &   --8601    &   $1029.53\pm  0.01$    & $0.37966\pm0.00001$   & $83\pm6$   &   $-115\pm5$     &  $0.052\pm0.002$    &  $4.3\pm0.5$    &   $7.8\pm0.9$ \\ 
\hline
\end{tabular}
\end{minipage}
\end{table*}  
A single Gaussian fit to the main component gives $1030.0450\pm0.0057$, i.e. $z =
0.378974\pm0.000008$ for the redshift of the absorber. This, in addition to the separate Gaussian
fits to the blue and redshifted features (see below), gives a velocity integrated optical depth of
$\int\, \tau\,dv = 17\pm3$ \kms. This is the strongest intervening 21-cm absorber yet found (see
table 1 of \citealt{ctm+07} and figure 12 of \citealt{ctd+09}), the next being at $z=0.52$ towards
B0235+164 with $\int\, \tau\,dv = 14$ \kms\ \citep{rbb+76}. The line strength gives a neutral
hydrogen column density of $N_{\rm HI} = 2.9\pm0.5\times10^{19}.\,(T_{\rm s}/f)$ \scm, where $T_{\rm
  s}$ is the spin temperature of the gas and $f$ the coverage of the background flux by the
absorber. Given that $T_{\rm s}/f\gapp100$~K is the minimum typical value of this degeneracy (see
\citealt{cmp+03,ctd+09}), it is clear that the neutral hydrogen column density in this absorber is
large, $N_{\rm HI}\gapp3\times10^{21}$ \scm, which is in the top 2\% of Sloan Digital Sky Survey
Data Release 5 DLAs \citep{pw09}.
The largest is $N_{\rm HI} = 8\times10^{21}$ \scm, a value which is exceeded by the absorber for 
$T_{\rm s}/f\gapp300$~K .




In addition to the main component, there is a strong red-shifted and a weaker blue-shifted component, 
 offset at $\approx+151$ and $\approx-143$ \kms, respectively.  These features have absorption line strengths which are
$\approx40$\% and $\approx20$\%
of the main profile ($\int\!\tau\,dv = 10\pm1$ \kms), the redshifted feature being as strong as any of those
in the main profile (Table \ref{fits}).  We
interpret these as being due to outflows from the nucleus of the galaxy, which contain a significant portion of
the absorbing gas. This is reminiscent of the Circinus galaxy, a near-by Seyfert in which the outflowing molecular gas mass is
comparable with that in the disk \citep{crjb98}. 

If the disk of the galaxy intervenes most of the background flux, as may be evident from the large optical depth, the
relatively narrow full-width half maximum of the main profile (FWHM$\,=86\pm6$ \kms) may suggest a low to intermediate
inclination for the galactic disk.  Thus, this may have an axis direction similar to that of the outflow, which we
believe is directed close to the line-of-sight, since it must intercept much of the flux from J0414+0534.\footnote{The diameter of
  the background emission is $5"$ at 1.4 GHz \citep{kmh97}, which is 26 kpc at $z=0.38$ ($H_{0}=71$~km~s$^{-1}$~Mpc$^{-1}$, $\Omega_{\rm matter}=0.27$ and
  $\Omega_{\Lambda}=0.73$).} However, this is not a
necessity, as the galactic disk (in which the absorption occurs) need not be coplanar with the circumnuclear torus, invoked by
unified schemes of active galactic nuclei, which collimates the outflow (\citealt{cw10} and references therein).

\subsection{The OH 18-cm absorption}
\label{oh}

\begin{figure*}
\centering \includegraphics[angle=270,scale=0.63]{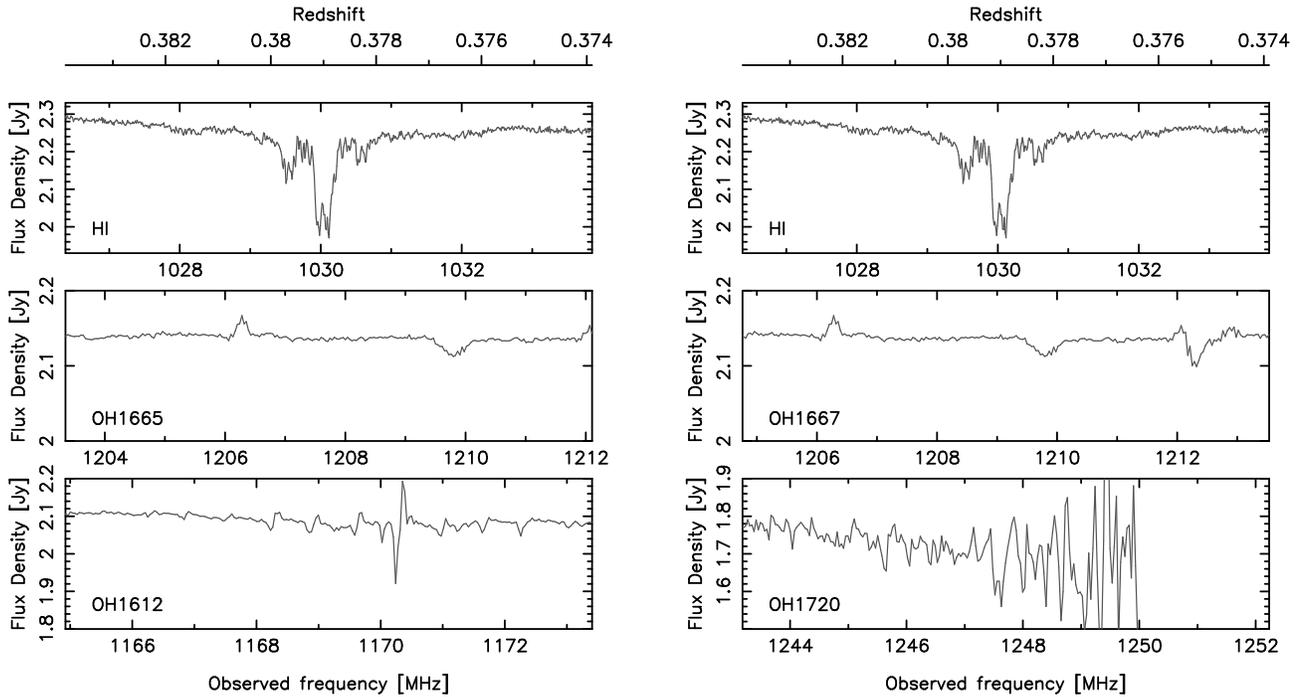} 
\caption{The OH  data at frequencies close to the redshift of the \HI\ absorption. Shown at 
a resolution of 10 \kms.}
\label{oh-spectra}
\end{figure*}

Since the purpose of our  spectral scans is to find the intervening molecular gas responsible for the reddening of the
background quasar, the frequencies covering the 
four 18~cm $^2\Pi_{3/2}$ OH lines were observed simultaneously in separate IFs. Given the redshift of
the 21-cm absorption, we expect the OH transitions to occur at 1169.15 (1612), 1207.71 (1665), 1209.13 (1667) and
1247.69 MHz (1720 MHz).  Examining the data which contain these frequencies (Fig. \ref{oh-spectra}), we see that there
may indeed be two features coincident with the 1665 ($F=1-1$) and 1667 MHz ($F=2-2$) transitions ({\sf OH1667} panel), although, even after
removal of the worst RFI, the bandpass is somewhat bumpy. From a Gaussian fit to the higher frequency feature, we obtain
a centroid of $1212.314\pm0.009$ MHz, which is where the $F=2-2$ transition would occur at a redshift of $z=0.37535\pm0.00001$
implying that the $F=1-1$ should be redshifted to $1210.89\pm0.01$ MHz. However, this is observed at $1209.80\pm 0.03$,
which in conjunction with the fact that $z=0.37535$ is out of the range of any of the \HI\ features (Table~\ref{fits}),
leads us to conclude that these are artifacts of an unstable bandpass.

\begin{figure}
\centering \includegraphics[angle=270,scale=0.75]{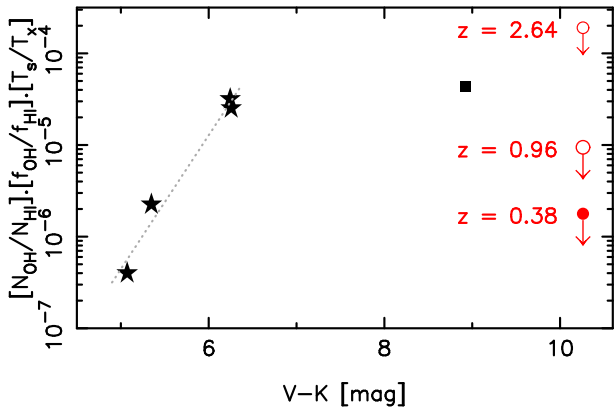}
\caption{The normalised OH line strength 
versus optical--near-IR colour for the five known OH absorbers and the searches towards J0414+0534. The stars
represent the four millimetre absorption systems and the square
represents PKS 0132--097 where OH, but no millimetre, absorption has been
detected (see main text). The line shows the least-squares fit to the millimetre absorbers. The filled circle shows the limit
presented here, and the unfilled circles the limits at the lens \citep{cdbw07} and the host redshifts (see main text).}
\label{OHoverH}
\end{figure}

Since the 1665 MHz band is the cleanest over the 21-cm redshift,  we use this transition to obtain an optical
depth limit. \citet{cdbw07} found a correlation between the \HI\ and OH profile widths for
the five known redshifted OH absorbers and so we resample the
r.m.s. noise level of 1.8 mJy per 10 \kms\ to the FWHM of
the main 21-cm absorption profile. This gives a $3\sigma$ limit of $\int\!\tau_{\rm 1665~MHz}\,dv \leq 0.074$ \kms\ per 86 \kms\ or
$N_{\rm OH} \leq3.3\times10^{13}\,(T_{\rm x}/f)$, where $T_{\rm x}$ is the excitation temperature of the gas.
Normalising this by the line strength of the main 21-cm profile gives $N_{\rm OH}/N_{\rm HI}\leq1.8\times10^{-6}\,.\,(f_{\rm HI}/f_{\rm OH})\,.\, (T_{\rm x}/T_{\rm s})$,
which is five times more sensitive than our previous limit in the $z=0.96$ gravitational lens (Fig. \ref{OHoverH}).

If the molecular abundance is correlated with the red colour of the background quasar \citep{cwm+10}),
it is apparent that none of the two known intervening 21-cm absorbing systems
towards J0414+0534 is the cause of the reddening. Furthermore, from OH observations at the host
galaxy redshift (Curran et al., in prep.), we have obtained a limit of $\int\!\tau_{\rm
  1612~MHz}\,dv \leq 0.16$ \kms\ per 5 \kms\ from the $F=1-2$ transition.\footnote{The redshifted frequencies of
  the other three transitions were completely dominated by RFI.} Rescaling this to the FWHM of
$\approx320$ \kms\ for the 21-cm profile \citep{mcm98}, gives $N_{\rm OH} \lapp3\times10^{14}\,(T_{\rm x}/f)$, or a
normalised line strength of $N_{\rm OH}/N_{\rm HI}\lapp2\times10^{-4}\,.\,(f_{\rm HI}/f_{\rm
  OH})\,.\, (T_{\rm x}/T_{\rm s})$, which may not rule out strong OH absorption in the host galaxy
(Fig. \ref{OHoverH}). Note that the low HCN abundance ($N_{\rm HCN}\leq1.2\times10^{13}$ cm$^{-2}$,
for $T_{\rm x}=10$~K) found by \citet{mcm98} may not rule out a large molecular abundance either, on
the basis that \citet{kcl+05} detect strong OH, but no HCO$^+$, absorption towards PKS 0132--097,
which \citet{cdbw07} suggest is due to differences in the coverage of the millimetre and
decimetre-wave emission. Given that three \HI\ absorption systems are now known towards this 
source, however, it is feasible that the red colour of the background quasar arises from an accumulation of
systems, rather than a single dusty intervening galaxy.

\subsection{The origin of the absorption}
\label{other}

\begin{figure}
\vspace{7.2cm}
\includegraphics{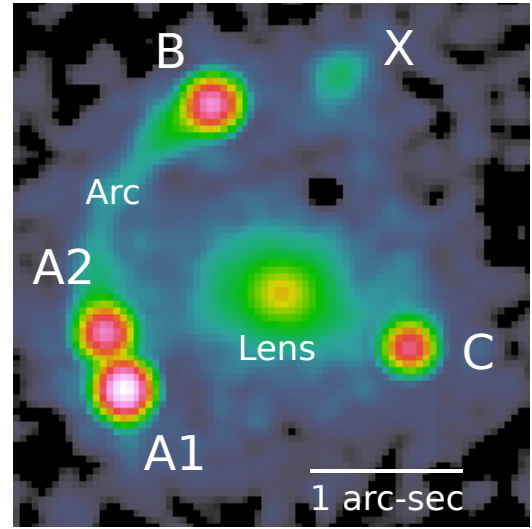}
\caption{$I$-band image of the lens and lensed components (A1, A2, B
\& C), together with Object X \citep{fls97}, from the  CASTLES  survey \citep{kfi+99}
[http://www.cfa.harvard.edu/castles/]. Resolved (22~GHz) radio emission is detected from
the lensed components (\citealt{kmh97}, see figure of 2 of \citealt{cdbw07}).}
\label{lens}
\end{figure}

Having discovered the absorption, the question of its origin arises. There have been a number of optical/near-infrared studies 
of J0414+0534 \citep{sm93,avhm94,fls97}, and these show, in addition to the four quasar images (A1, A2, B
\& C) and the lensing galaxy, a feature often referred to as ``Object X'' (Fig. \ref{lens}).
It is located about 1 arc-sec west of component 'B', giving impact parameters of between 5--15~kpc
(for  $z=0.38$) and 8--25 kpc (for  $z=0.96$) to the four quasar images. 
HST photometry \citep{fls97} gives $R>26.268\pm0.0.063$ and $I=24.769\pm0.063$,
thus having a different colour to the quasar images, while probably being slightly redder than the lensing galaxy.

We note that the spectrum of \citet{tk99} shows tantalising evidence of peaks approximately where the \oiii\ $\lambda\lambda4959,5007$\AA\ doublet
would be located for the \HI\ absorption redshift of $z=0.3789$ (observed wavelengths of $6837$ and $6903$\AA). 
The slit position, as indicated in their figure 1, would indeed lie directly across Object X and 
large  \oiii\  emission line fluxes would be expected from an AGN or gas-rich galaxy undergoing rapid star-formation,
either of which are consistent with the observation of rapid outflows of \HI\ in each direction.

\section{Summary}

As part of an ongoing project, scanning the entire redshift space towards highly reddened radio sources,
in search for the object responsible for the obscuration of the optical light, we have detected
a second intervening 21-cm absorber at $z=0.38$ towards the $z= 2.64$ quasar MG J0414+0534.
The other intervening absorber arises in the $z=0.96$  gravitational lens \citep{cdbw07} and,
combined with the  absorption found at the host redshift \citep{mcm98},  gives a total
of three \HI\ absorbers so far detected along this sight-line.

Although we cannot determine the $T_{\rm s}/f$ degeneracy for any of the three 21-cm absorbers, thus deriving
the column densities, the 21-cm line strength in this new absorber is by far the strongest detected along
this sight-line, being four times stronger
than in the host galaxy [$N_{\rm HI}=7.5\pm1.3\times10^{18}.\,(T_{\rm s}/f)$ \scm] and 19 times stronger than
 in the lens [$N_{\rm HI}=1.6 \times 10^{18}\,(T_{\rm s}/f)$ \scm]. In fact, the strongest intervening
21-cm absorber found to date.

Despite the \HI\ absorption strength and the very red colour of this source ($V-K = 10.26$), OH remains undetected
to very strong limits [$N_{\rm OH}/N_{\rm HI}\leq1.8\times10^{-6}\,.\,(f_{\rm HI}/f_{\rm OH})\,.\, (T_{\rm x}/T_{\rm s})$],
inferring that this new absorber is not the primary cause of the red colour. OH is also undetected in the lens and the host
galaxy, although the latter is to relatively weak limits not allowing us to rule out that this is where much of the reddening occurs.

We suggest that the absorption may be associated with the feature known as Object X in the optical field of J0414+0534:
This could be spatially coincident with the spectrum of \citet{tk99}, of which the $6837$ and $6903$\AA\ features are
consistent with the expected wavelengths for the \oiii\ doublet at the redshift of the 21-cm absorption feature.
Not being a companion of the lensing galaxy would have implications for the current lens models (e.g. \citealt{twh00,kd04}). 

This is the fourth only of the (now) 42 known redshifted intervening 21-cm absorbers, which have been {\em discovered} through
21-cm absorption, the others being at $z= 0.44$ towards B0809+483 \citep{bm83}, $z = 0.69$ towards B1328+307
\citep{br73} and $z= 0.78$ towards B2351+456 \citep{dgh+04}. The vast majority have been found by tuning the receiver to
a frequency based upon the redshift of an a priori detected absorber (Mg{\sc ii} or
Lyman-\AL). Being optically selected, this traditional method biases against the more dust obscured systems and the discovery of two
intervening 21-cm absorbers from spectral scans towards J0414+0534, suggests that future surveys with Square
Kilometre array may uncover a large population of faint, dusty high redshift galaxies. 

\section*{Acknowledgements}
We would like to thank Jeremy Darling who provided Fig.~\ref{lens}.
This research has made use of the NASA/IPAC Extragalactic Database
(NED) which is operated by the Jet Propulsion Laboratory, California
Institute of Technology, under contract with the National Aeronautics
and Space Administration. This research has also made use of NASA's
Astrophysics Data System Bibliographic Services. 

The National Radio Astronomy Observatory is a facility of the National
  Science Foundation operated under cooperative agreement by Associated Universities, Inc.


\label{lastpage}

\end{document}